\documentclass[12pt]{article}
\usepackage{amsmath,epsf,amssymb,latexsym,cite}

\makeatletter
\renewcommand\section{\@startsection {section}{1}{\z@}%
                                   {-3.5ex \@plus -1ex \@minus -.2ex}%
                                   {2.3ex \@plus.2ex}%
                                   {\normalfont\bfseries}}
\makeatother

\newif\iffigs\figstrue

\DeclareFontFamily{U}{rsf}{}
\DeclareFontShape{U}{rsf}{m}{n}{
  <5> <6> rsfs5 <7> <8> <9> rsfs7 <10-> rsfs10}{}
\DeclareMathAlphabet\Scr{U}{rsf}{m}{n}

\def\pplogo{\vbox{\kern-\headheight\kern -29pt
\halign{##&##\hfil\cr&{
\ppnumber}\cr\rule{0pt}{2.5ex}&\ppdate\cr}
}}
\makeatletter
\def\ps@firstpage{\ps@empty \def\@oddhead{\hss\pplogo}%
  \let\@evenhead\@oddhead 
}
\def\maketitle{\par
 \begingroup
 \def\thefootnote{\fnsymbol{footnote}}
 \def\@makefnmark{\hbox{$^{\@thefnmark}$\hss}}
 \if@twocolumn
 \twocolumn[\@maketitle]
 \else \newpage
 \global\@topnum\z@ \@maketitle \fi\thispagestyle{firstpage}\@thanks
 \endgroup
 \setcounter{footnote}{0}
 \let\maketitle\relax
 \let\@maketitle\relax
 \gdef\@thanks{}\gdef\@author{}\gdef\@title{}\let\thanks\relax}
\makeatother

\def\O{\Scr{O}}
\def\C{{\mathbb C}}
\def\P{{\mathbb P}}

\def\Z{{\mathbb Z}}

\def\Hom{\operatorname{Hom}}

\def\CY{Calabi--Yau}
\def\LG{Landau--Ginzburg}

\def\cA{{\Scr A}}

\def\cF{{\Scr F}}

\def\DC{\mathbf{D}}

\def\labto#1{\mathrel{\mathop\to^{#1}}}

\begin{document}
\setcounter{page}0
\def\ppnumber{\vbox{\baselineskip14pt
\hbox{DUKE-CGTP-03-07}
\hbox{hep-th/0312188}}}
\def\ppdate{December 2003} \date{}

\title{\LARGE The Breakdown of Topology at Small Scales\\[10mm]}
\author{
Paul S.~Aspinwall\\[2mm]
\normalsize Center for Geometry and Theoretical Physics \\
\normalsize Box 90318 \\
\normalsize Duke University \\
\normalsize Durham, NC 27708-0318
}

{\hfuzz=10cm\maketitle}

\def\Large{\large}
\def\LARGE{\large\bf}

\vskip 1cm

\begin{abstract}
We discuss how a topology (the Zariski topology) on a space can appear
to break down at small distances due to D-brane decay. The mechanism
proposed coincides perfectly with the phase picture of \CY\ moduli
spaces. The topology breaks down as one approaches non-geometric
phases. This picture is not without its limitations, which are also discussed.
\end{abstract}

\vfil\break


\section{Introduction}    \label{s:intro}

The concept of spacetime is supposed to be derived rather than
fundamental in superstring theory. Many ideas have emerged concerning
how one might obtain such a derivation but none seems to be without
its flaws.

To avoid the extraordinary complications introduced by the presence of
a nontrivial time-like direction, it is simplest to study a
compactification model. Here one can ask how to deduce the geometry of
a \CY\ space $X$ of complex dimension $d$ used to compactify string
theory down to $10-2d$ dimensions. Then one can ask to what extent
the geometry of $X$ can be deduced from ``intrinsic'' properties of the
string, which in many cases can be paraphrased in terms of the physics
of the uncompactified dimensions.

It is a repeated phenomenon in string theory that questions associated
to compactifications tend to be most mathematically interesting in
the case $d=3$ which happily coincides with observed 4-dimensional
spacetime. It is this case that we consider here.

Let us consider a type II string compactification. The D-branes
corresponding to BPS solitons of the resulting $N=2$ theory are the focus of
our attention. Following the work of Kontsevich \cite{Kon:mir} and
Douglas \cite{Doug:DC}, it is known that some of the D-branes (the
``B-branes'') are described by the derived category of coherent
sheaves on $X$. In this context, the conditions for stability
\cite{DFR:stab,AD:Dstab,Brg:stab} are now understood to a fair
extent. In particular, it is known that many D-branes which are
understood to exist in flat space become unstable as the
characteristic radius of a \CY\ becomes small with respect to the
scale set by the string tension.

In this short paper we outline a simple idea for how one might use
such a decay of D-branes to motivate the notion that topology is only
well-defined in the large radius limit. We should immediately concede
that this argument is not without its limitations not least of which
is our use of the Zariski topology. This topology is not usually associated
with physics. The main point we wish to emphasize is that the picture
we propose coincides beautifully with the ``phase'' picture of
\cite{W:phase,AGM:II}. At least in the context of linear
$\sigma$-models or, equivalently, toric geometry, the moduli space of
$N=(2,2)$ superconformal field theories can be roughly divided into
phases, each with a particular interpretation. The ``geometric phases''
consist of the smooth \CY\ $X$ together with other phases representing
singular versions of $X$ such as orbifolds. The ``non-geometric
phases'' consist of interpretations such as \LG\ models, fibrations
with a \LG\ fibre (``hybrid models'') and exoflops which have a \LG\
model associated to some subspace. In the language of algebraic
geometry, these \LG\ models can be associated with non-reduced
schemes.

We will see that the topology is lost as one moves from the geometric
phases to the non-geometric phases as one would hope.  In
\cite{me:navi,AL:DC} the author hoped that 0-branes might play the key
r\^ole in this loss of geometry or topology. Indeed, it is 0-branes
that decay as one passes through a flop \cite{me:point}.  However, in
non-geometrical phases the 0-brane remains stubbornly stable
\cite{AD:Dstab,Schd:D0}. Instead, as realized in
\cite{Doug:DC,Doug:S01}, it is the 4-branes which naturally decay
along the phase boundaries. Our purpose here is to argue that this can
be associated with a loss in topology and that it is a general feature
of all models associated to toric geometry. So the picture appears to
be that 0-brane decay is associated with a ``change'' in topology
while 4-brane decay is associated to a ``loss'' of topology.


\section{The Zariski Topology}    
     \label{s:con}

Let $X$ be a topological space. That means that $X$ is a set of points
and we have a set of subsets of $X$, called ``open sets'', satisfying
the following conditions:
\begin{enumerate}
\item $X$ itself and the empty set $\emptyset$ are both open.
\item Any union of open sets is open.
\item Any finite intersection of open sets is open.
\end{enumerate}

If $X$ has a metric, then one can form a topology with open sets
consisting of unions of interiors of balls of any radius $\epsilon>0$
and any center. This ``metric topology'' is probably the one we are
most used to thinking about in physics.

In algebraic geometry, the natural topology is the ``Zariski
topology'' defined as follows. Let $X$ be embedded in $\P^N$ as the
intersection of the zeroes of some algebraic equations in the
homogeneous coordinates of $\P^N$. If $f$ is any algebraic equation in
the homogeneous coordinates of $\P^N$, then the intersection of $f=0$
with $X$ will be an ``algebraic'' subspace of $X$ (typically of
codimension one). Let $U_f$ be the {\em complement\/} of this subspace
in $X$.

The Zariski topology on $X$ is then defined as the open sets
consisting of $\emptyset$, $X$, unions of $U_f$'s, and finite
intersection of $U_f$'s, for any $f$'s. In other words, the open sets
consist of complements of algebraic subspaces of $X$.

Note that the Zariski topology can be built from a
knowledge of the algebraic codimension one subspaces, i.e., {\em
divisors\/} of $X$.

It should be admitted from the outset that the Zariski topology is
peculiar by physicists standards. For example, it is never Hausdorff
in nontrivial examples. The problem is that it is very ``coarse'',
i.e., it consists of relatively few open sets. Having said that, for
some purposes (e.g., computing cohomology) it is equivalent to a
metric topology. This relationship was explored in the famous GAGA
paper \cite{Ser:GAGA}.

Knowing to what extent the Zariski topology suffices for string theory
is probably a deep question and we will certainly not attempt to
answer it here. Let us just suggest that many approaches to string
theory are algebraic in nature, and in such a context one should
expect something like the Zariski topology to be very natural.


\section{D-Branes}    
     \label{s:dbn}

The Zariski topology suggests that we are interested in the following
D-branes on the \CY\ threefold $X$:
\begin{enumerate}
\item The 0-branes which form the ``points''.
\item The 4-branes which form the divisors.
\end{enumerate}
These are all B-branes. That is, they are BPS branes which are preserved
by the twist to the topological B-model. 

In this paper we will restrict our attention to the zero
string-coupling limit.  In this case it is established that
B-branes are described by the derived category of coherent sheaves on
$X$ \cite{Kon:mir,Doug:DC,Laz:DC,AL:DC,Dia:DC,me:tasi03}.

The 0-branes
consist of (a complex with a single nonzero element given by) a
skyscraper sheaf $\O_p$ for $p\in X$. Similarly the 4-branes are
associated to the sheaves $\O_D$ supported on divisors $D\subset X$.

It is an interesting question to ask to what extent one might determine
the set of 0-branes given only the purely categorical (i.e.,
worldsheet) description of the D-branes. This was discussed in the
context of string theory in \cite{me:point} based on the work of
Bondal and Orlov \cite{BO:DCeq}. Here we will assume that a
determination of the 0-branes and 4-branes has been made.

Given a 0-brane corresponding to $p$ and a 4-brane corresponding to
$D$, there is an open string between these D-branes in the topological
sector precisely when $p\in D$. Thus the topological field theory has
enough information to define the Zariski topology. In derived category
language, the open set $U_D$, equal to the complement of $D$, is given
by the set
\begin{equation}
  U_D = \{ \O_p\, |\, \Hom(\O_D,\O_p)=0 \}.
\end{equation}

It was realized by Douglas \cite{Doug:DC,Doug:S01} that the 4-branes
can decay as one moves away from the large radius limit. It is this
observation that we will exploit here. 


\section{The Quintic Threefold}    
     \label{s:quin}

Let us quickly review a result from \cite{AD:Dstab}.
Suppose $X$ is the quintic threefold and is given by the embedding
$i:X\hookrightarrow\P^4$. 

Let $\O(k)$ denote the pullback, via $i$, of the usual twisted sheaves
on $\P^4$. We then have a short exact sequence
\begin{equation}
  0\to \O(-1) \labto f \O \to \O_D\to0,
\end{equation}
where $D$ is a degree one divisor in $X$ associated to the linear map
$f$.

\iffigs
\begin{figure}[ht]
  \centerline{\epsfxsize=15cm\epsfbox{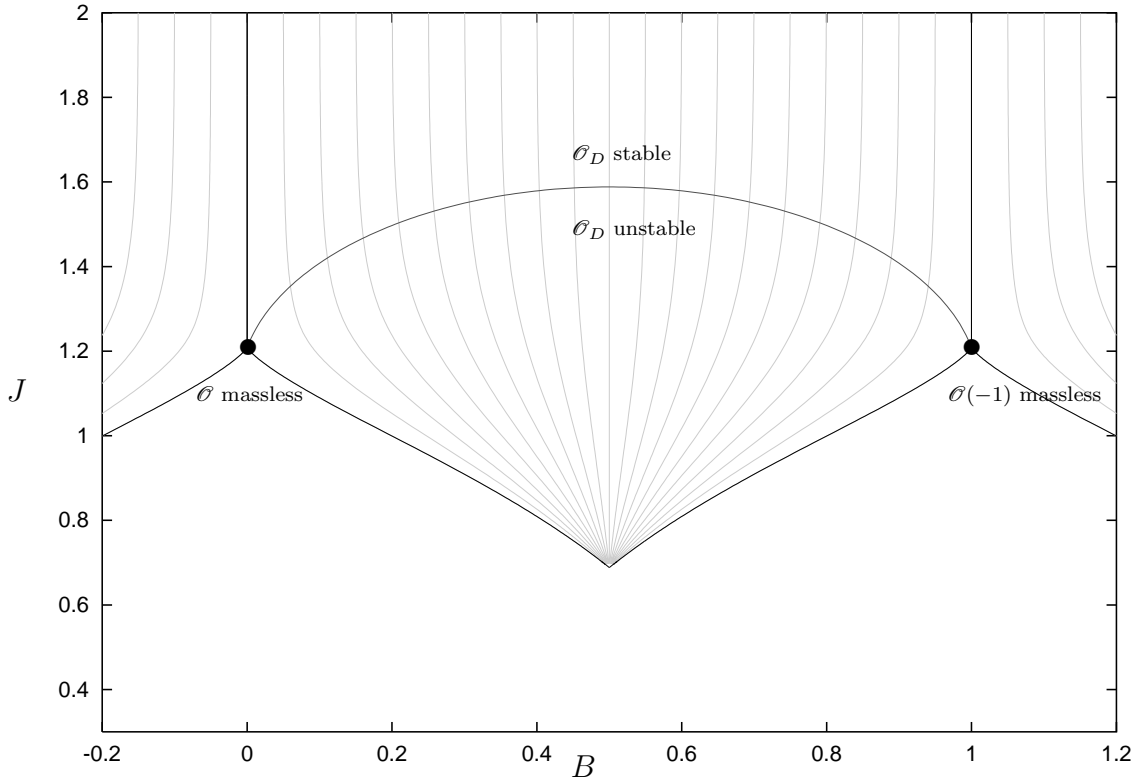}}
  \begin{picture}(0,0)(0,0)
    \put(37.1,62.1){\circle*{2}}
    \put(30,55){\text{\scriptsize$\O$ massless}}
    \put(133.3,62.1){\circle*{2}}
    \put(130,55){\text{\scriptsize$\O(-1)$ massless}}
    \put(80,87){\text{\scriptsize$\O_D$ stable}}
    \put(80,77){\text{\scriptsize$\O_D$ unstable}}
    \put(80,5){\text{$B$}}
    \put(5,55){\text{$J$}}
  \end{picture}
\vspace{-0.7cm}
  \caption{4-brane decay on the quintic.}
  \label{f:q1}
\end{figure}
\fi

Ignoring other possible decays, the stability of $\O_D$ is determined
by the mass of the open string associated with the map $f$. $\O_D$ is
stable if and only if this open string is tachyonic. This mass was
computed explicitly in \cite{AD:Dstab} and the result is shown in
figure~\ref{f:q1}. This figure shows the complex $B+iJ$ plane. 
This plane is divided into non-tesselating fundamental regions of the
moduli space as discussed in \cite{CDGP:}. At
large radius $\O_D$ is stable. As we cross the line of marginal
stability the 4-brane decays into $\O$ and $\overline{\O(-1)}$.

The line of marginal stability ends at the points where the decay
products become massless. This is due to a general rule of
$\Pi$-stability as argued in \cite{AD:Dstab} --- a line of marginal
stability will either end at a point where a decay product becomes
massless or it will end on another line of marginal stability for that
decay product.

The decay products $\O$ and $\O(-1)$ become massless for $B=0$ and
$B=1$ respectively. The line of marginal stability forms a
``wall'' across the fundamental region shown in the figure. This ties
in with the ``phase'' picture for $N=(2,2)$ superconformal field
theories studied in \cite{W:phase,AGM:II}. The 4-brane is stable while
we are in the ``\CY\ phase'' and becomes unstable as we pass into the
``Landau-Ginzburg'' phase.

So far we only considered divisors of degree one. One may also
consider higher degree cases given by the exact sequence:
\begin{equation}
  0\to \O(-k) \labto f \O \to \O_{D_k}\to0,
\end{equation}
where $D_k$ is a divisor of degree $k$. As discussed in
\cite{AD:Dstab}, such a 4-brane decays in a similar way to the degree
one case but will decay at a larger radius depending on $k$.

The picture we wish to present is then as follows. Near large radius
limit, very nearly all of the 4-branes on $X$ are stable, allowing us to
determine the Zariski topology to a good approximation. Then as the
radius of $X$ decreases, we start to lose 4-branes of progressively
lower degree. As we pass out of the \CY\ phase, we lose the last
4-brane and the Zariski topology is completely lost.


\section{The General Phase Picture}    
     \label{s:phas}

A general \CY\ threefold $X$ will have more than one deformation of
$B+iJ$ and so the above picture of walls of marginal stability will be
become considerably more complicated. Fortunately we know enough about
general features of the moduli space to generalize the above
computation.

Suppose that $H^2(X,\Z)$ is generated by $e_j$, where $j=1\ldots
h^{1,1}(X)$. Suppose further that each $e_j$ is positive in the sense
that it is Poincar\'e dual to an algebraic 4-cycle $D_j\subset
X$. These divisors $D_j$ are the 4-branes whose decay we wish to
analyze. 

Assume the decay is associated to the obvious sequence:
\begin{equation}
  0\to \O(-D_j) \labto f \O \to \O_{D_j}\to0,
\end{equation}
where $\O(-D_j)$ is the line bundle on $X$ with $c_1=-e_j$.
This implies that the wall of marginal stability will stretch between
a locus where the basic 6-brane $\O$ becomes massless and a locus
where $\O(-D_j)$ becomes massless. When these branes become
massless, the associated conformal field theory becomes singular
and we are on the discriminant locus within the moduli space of
$B+iJ$. This is equivalent to the classical discriminant in the moduli
space of complex structures of the mirror to $X$.

Except for the simplest of examples, an explicit computation of the
discriminant is not practical. Suppose we restrict attention to the
large ``Batyrev--Borisov'' class of complete intersections in a toric
variety \cite{Bat:m,Boris:m}. Virtually all known \CY\ threefolds fall
into this class. Many features of the discriminant in this case have
been studied by Gelfand, Kapronov, and Zelevinsky
\cite{GKZ:book}. Much of this structure has also been analyzed
directly from the linear $\sigma$-model by Morrison and Plesser
\cite{MP:inst}.

The \CY\ $X$ is associated to a set of points $\cA$ on a lattice $N$
(see \cite{AG:gmi}, for example, for more details). Let $P_X$ be the
convex polytope of this point set. The discriminant is reducible with
each irreducible component associated with a face (of any codimension)
of this polytope. One irreducible component is distinguished ---
namely the one associated to $P_X$ itself. We call this the ``primary
component'' of the discriminant which we denote $\Delta_0$.

It has been conjectured \cite{Kont:mon,Mor:geom2,Horj:DX,me:navi} that
this primary component of the discriminant is precisely where the
basic 6-brane $\O$ becomes massless. This is a very natural conjecture
and has been shown to be true in several examples. We will assume it
to be true.

The statement that a certain D-brane becomes massless for a certain
component of the discriminant locus is not really well-defined. One
must specify some basepoint in the moduli space (usually near large
radius) to define the labeling of the set of D-branes and then specify
the path taken to the discriminant. It follows that the above
conjecture should really assert that, for some naturally defined short
path from the large radius limit to the primary component of the
discriminant, the D-brane $\O$ becomes massless.

Suppose we loop around the large radius limit point before we embark
on this path towards the primary component $\Delta_0$. Under the
established rules (see, for example, \cite{Horj:DX,AD:Dstab}) the
monodromy around large radius limit gives an autoequivalence of
$\DC(X)$ given by $\cF\mapsto\cF\otimes L$ for some line bundle $L$.
$c_1(L)$ is determined by exactly how one went around the large radius
limit. In particular, if we perform this monodromy by varying the
component of $B+iJ$ associated to $e_j$, we obtain $L=\O(-D_j)$. Thus
the D-brane $\O(-D_j)$ also has vanishing mass on $\Delta_0$ --- we
just have to loop around the large radius limit in the right way
before heading off towards $\Delta_0$.

The case of the quintic in figure~\ref{f:q1} should help the reader
visualize this construction. Here the primary component of the
discriminant consists simply of the ``conifold point''. The D-brane
$\O$ indeed becomes massless here but $\O(-1)$ becomes massless at an
image of this point in the Teichm\"uller space under monodromy around
the large radius limit.

We have therefore established a higher-dimensional analogue of
figure~\ref{f:q1}. The wall of marginal stability will form a wall
which encloses the large radius phase in the moduli space of $B+iJ$ in
the direction of a given $e_j$. By considering each divisor $D_j$ in
turn we may completely wall off the large radius limit.  The location
of this (real co-dimensional one) wall is guided by the location of
the (complex co-dimensional one) primary component of the discriminant
locus $\Delta_0$. Each one-complex-dimensional slice of the moduli
space should look roughly like figure~\ref{f:q1}.

\iffigs
\begin{figure}
  \centerline{\epsfxsize=10cm\epsfbox{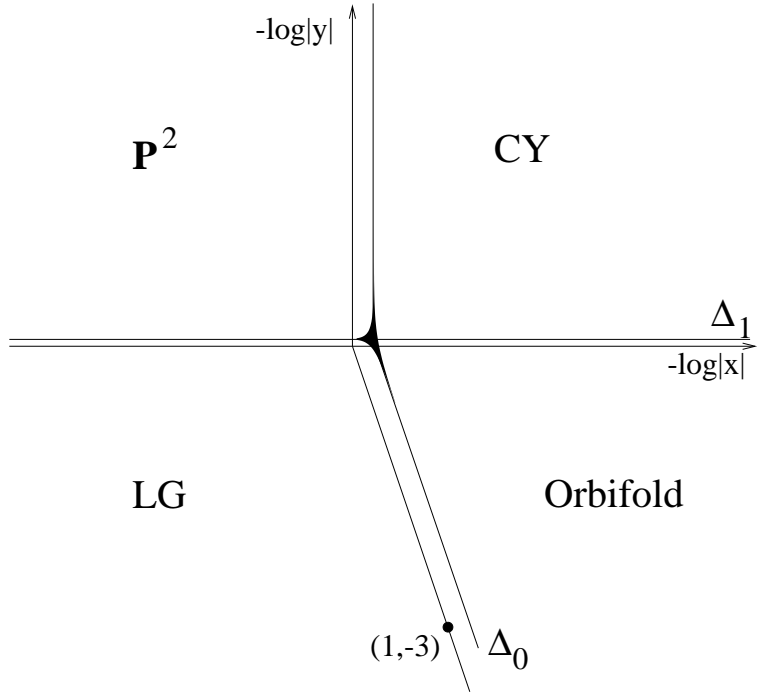}} 
  \caption{The 4 phases of the 2 parameter example.}
  \label{f:phase}
\end{figure}
\fi

To understand a more complicated example consider the much-studied
two-parameter model of \cite{CDFKM:I}. Our notation is taken from
\cite{me:navi}. Here we have four phases consisting of a \CY, an
orbifold, a hybrid and a Landau--Ginzburg theory. The hybrid consists
of a fibration over $\P^2$ with fibre given by a Landau--Ginzburg
theory. Projecting this moduli space into the ``algebraic'' $J$-plane
as discussed in \cite{AGM:sd}, we obtain figure~\ref{f:phase}.
The discriminant locus has two components shown as $\Delta_0$ and
$\Delta_1$ in the figure. The primary component, $\Delta_0$, lies in 3
of the 4 walls forming the phase ``boundaries''. The wall between the
\CY\ phase and the orbifold phase contains only $\Delta_1$.

Imagine passing from the \CY\ phase to the hybrid phase. Since we
cross $\Delta_0$, all the 4-branes dual to that direction in $H^2(X)$
decay, thus destroying the Zariski topology.\footnote{The 4-branes
wrapping the $\P^2$ base are still stable. One might use this to say
that part of this model still has a geometrical interpretation as
would befit a ``hybrid'' model.} Passing from the \CY\
phase to the orbifold phase involves blowing down an exceptional
$\P^2$. This does {\em not\/} involve crossing $\Delta_0$ and so we do not
expect a loss of topology. Indeed, the 4-branes which wrap the
exceptional $\P^2$, which are the ones one might expect to decay,
instead become massless on $\Delta_1$. Proceeding further to the
Landau--Ginzburg phase should cause theses 4-branes to decay.

We therefore have a Zariski topology apparently well-defined in the
\CY\ and orbifold limits of the theory, but not in the hybrid and
Landau--Ginzburg limits. That is, we have a topology in the
``geometric'' phases as one might expect.

One can define exactly what one means by ``geometric phases'' in the case
of hypersurfaces in toric varieties by demanding that each simplex in
the triangulation of $\cA$ has the unique interior point as a vertex
\cite{AGM:II}. A generalization to complete intersections exists.
As argued in \cite{me:navi}, based on the results in \cite{GKZ:book},
the phase boundaries between geometric and non-geometric phases always
contain $\Delta_0$ whereas the phase boundaries between two geometric
phases never contain $\Delta_0$. Therefore, {\em we always lose the
Zariski topology as we leave the geometric phases.}


\section{Problems}   \label{s:fly}

The reader may have noticed that we have an unnecessarily restrictive
definition of a 4-brane. According to the usual correspondence between
sheaves and bundles, the sheaf $\O_D$ corresponds to the trivial line
bundle over $D$ (extended by zero to form a sheaf over
$X$).\footnote{This correspondence actually gets shifted in string
theory. See \cite{KS:Ext} for more details.}  Why not choose a
nontrivial line bundle over $D$?

For example, in the quintic there is the 4-brane $\O_D(1)$ defined by
the sequence:
\begin{equation}
  0\to \O \labto f \O(1) \to \O_D(1)\to0.
\end{equation}
The mode of decay of $\O_D(1)$ represented by this sequence forms a
line of marginal stability between the points where $\O$ and $\O(1)$
become massless. In terms of figure~\ref{f:q1}, this would move the
line of marginal stability by a shift of $B$ by one unit to the left.
Thus we {\em could\/} make this 4-brane D-decay by going around the
large radius limit once before heading off to the \LG\ phase. However,
if we head straight for the \LG\ phase this decay will not happen.

It would be nice if we could show that all such 4-branes really decay
by some other channel. The current understanding of $\Pi$-stability
makes it very difficult to assess the stability of any D-brane except
in the easiest of examples. Several obvious potential decay modes for
$\O_D(1)$ were checked without success. However, given the lack of a
systematic method for checking decay modes, this does not constitute
evidence that the D-brane is stable.

One might try to find the stable D-branes at the Gepner point directly
using boundary conformal field theory methods such as in
\cite{RS:DGep,BDLR:Dq,FSW:DGep}. It appears to be a formidable task to
completely describe all the possibilities but there appears to be no
stable state with the charge of a 4-brane (with any number of 2-branes
added) of the limited types of D-brane studied in
\cite{RS:DGep,BDLR:Dq}.

An observation which might count against the decay of $\O_D(1)$ comes
from the analysis of the $\C^3/\Z_3$ orbifold. Let $E\cong\P^2$ be the
exceptional divisor in this case and let $C\cong\P^1$ be a complex
line on $E$. The 2-brane $\O_C$ wrapping $C$ can be shown to decay as
one moves from the large-blow-up limit to the orbifold point in a very
similar way to the decay of the 4-branes above
\cite{DFR:orbifold,me:tasi03}. However, the D-brane $\O_C(1)$ can be
shown to be stable in the neighbourhood of the orbifold point
\cite{Douglas:Dlect,me:tasi03}. Having said that, the rules for
orbifolds are not quite the same as for Gepner models. The analysis in
the orbifold case is facilitated by the central charges of the
B-branes lining up to have the same phase at the orbifold --- allowing
for quiver methods to be used \cite{Douglas:Dlect,me:tasi03}. This
does not happen for Gepner points.

If one can show that all the 4-branes decay on moving out of the
geometric phases, then it brings this work into line with the analysis
of Bondal and Orlov \cite{BO:DCeq}. They showed how the derived
category (i.e., D-branes) could be used to reconstruct an algebraic
variety complete with its Zariski topology by using such 4-branes.

On the other hand, if these 4-branes corresponding to nontrivial
bundles do tend to remain stable in the non-geometric phases, then we
need to motivate why it is the trivial line bundles that are essential
in building the Zariski topology.

One might take a completely different approach to determining the
target space from the BPS D-brane data. For example, one might say
that the target space is given by the moduli space of the
0-brane. Since the 0-brane is stable in the non-geometric phases, one
is free to do this. It is not obvious however that this is a perfect
definition of spacetime. For example, in \cite{DGM:Dorb} it was shown
that the metric seen by a 0-brane on an orbifold resolution is not the
expected one. I would like to argue that such an approach is essentially
different to the algebraically-motivated ideas in this paper. Such a
D-brane probe never sees the phase structure anyway since the
non-geometric phases get squeezed away as pictured in
\cite{AGM:sd,DGM:Dorb,W:MF}.

If one is wedded to such an idea of a ``differential'' approach to
spacetime geometry, then the ideas in this paper probably hold little
appeal. On the other hand, the more ``algebraic'' approaches to
spacetime should always require arguments about stability, such as the
ones in this paper, to organize ideas concerning topology.


\section*{Acknowledgments}

I wish to thank A.~Lawrence, I. Melnikov, D.~Morrison, R.~Plesser,
L.~Susskind and W.~Taylor for useful conversations. The author is
supported in part by NSF grants DMS-0074072 and DMS-0301476.


\end{document}